# Physical properties of Thallium-Tellurium based thermoelectric compounds using first-principles simulations

Xiaoma Tao[a,b], Philippe Jund[b], Romain Viennois[b] and Jean-Claude Tédenac[b]

[a] Department of Physics, Guangxi University, Nanning, 530004, P. R. China

[b] Institut Charles Gerhardt, Universite Montpellier 2, Pl. E. Bataillon CC1503 34095 Montpellier, France

Abstract:

We present a study of the thermodynamic and physical properties of $Tl_5Te_3$, $BiTl_9Te_6$ and $SbTl_9Te_6$ compounds by means of density functional theory based calculations. The optimized lattice constants of the compounds are in good agreement with the experimental data. The electronic density of states and band structures are calculated to understand the bonding mechanism in the three compounds. The indirect band gap of $BiTl_9Te_6$ and $SbTl_9Te_6$ compounds are found to be equal to 0.256 eV and 0.374 eV, respectively. The spin-orbit coupling has important effects on the electronic structure of the two semiconducting compounds and should therefore be included for a good numerical description of these materials. The elastic constants of the three compounds have been calculated, and the bulk modulus, shear modulus, and young's modulus have been determined. The change from ductile to brittle behavior after Sb or Bi alloying is related to the change of the electronic properties. Finally, the Debye temperature, longitudinal, transverse and average sound velocities have been obtained.

## 1. Introduction

Recently thermoelectric materials have attracted the interest of many researchers and a lot of investigations have been performed showing that high efficiencies could indeed be obtained[1]. Among the III-VI group, the Tl-Te system is of particular interest and has been the topic of many investigations for its thermoelectric properties[2-7]. This system exists in four crystalline phases: $Tl_2Te_3$, TlTe, $Tl_5Te_3$, and $Tl_2Te$[8]. The $Tl_2Te_3$ phase is semiconducting with an energy gap of 0.68±0.03 eV[3] and TlTe is a semimetal[2-4]. $Tl_5Te_3$ has a metallic behavior and exhibits a superconducting transition at low temperature[2,3,5] (the electronic structure and superconductivity of $Tl_5Te_3$ have been studied by Nordell and Miller[5]). The $Tl_5Te_3$ structure has four crystallographic sites: Tl1 site 16l; Tl2 site 4c; Te1 site 4a and Te2 site 8h. This is a $Cr_5B_3$-prototype body centered tetragonal structure with the space group I4/mcm (140)[9]. According to B. Wölfing et al[10,11], the three Te atoms in $Tl_5Te_3$ are in the valence state $Te^{2-}$, therefore the five Tl atoms have to provide six electrons. This is achieved by expanding the formula unit to $Tl^{1+}_9Tl^{3+}Te^{2-}_6$, where the 4c site, which accommodates two Tl atoms per expanded formula unit, is equally occupied by $Tl^+$ and $Tl^{3+}$. Thus, the $Tl^{3+}$ can be substituted with the trivalent elements $Bi^{3+}$ and $Sb^{3+}$, resulting in the compounds $BiTl_9Te_6$ and $SbTl_9Te_6$. Some studies have already been performed on $BiTl_9Te_6$[10-12] and $SbTl_9Te_6$[11] due to their excellent thermoelectric properties and this explains why these compounds are the topic of the present study. Nevertheless thallium is a toxic element which can not be used directly in thermoelectric materials. But a fundamental study of this material can be done to understand the properties of other kind of materials crystallizing in the same structure.

Even though the knowledge of the different physical properties of these compounds is important, there are only few studies focused on the electronic and mechanical properties of $Tl_5Te_3$, $BiTl_9Te_6$ and $SbTl_9Te_6$. In this work, the thermodynamic, electronic, thermoelectric and elastic properties of $Tl_5Te_3$, $BiTl_9Te_6$ and $SbTl_9Te_6$ compounds with $Cr_5B_3$-type structure have been studied using first-principles calculations based on the Density Functional Theory (DFT).

The remainder of this paper is organized as follows. In Section 2, the method and the calculation details are described. In section 3, the formation enthalpy, the electronic, thermoelectric and structural properties as well as the elastic constants and the Debye temperature are presented and discussed. Finally, some conclusions are drawn in Section 4.

## 2. Computational details

First-principles calculations are performed by using the projector augmented-wave (PAW) method[13, 14] within the generalized gradient approximation (GGA), as implemented in the highly-efficient Vienna Ab initio Simulation Package (VASP)[15, 16]. The calculations employed the Perdew-Bucke-Ernzerhof (PBE)[17] exchange-correlation functional in the GGA. For the PAW calculations we use potentials with 15 valence electrons for Bi ($5d^{10}6s^26p^3$), 5 for Sb($5s^25p^3$), 13 for Tl ($5d^{10}6s^26p^1$) and 4 for Te($5s^25p^4$), and a plane-wave energy cutoff of 300 eV is held constant for all the calculations. Brillouin zone integrations are performed using Monkhorst-Pack k-point meshes[18], with a k-point sampling of 11x11x11 for the three compounds and the tetrahedron method with Blochl correction[19] is used in the present calculations. The Methfessel-Paxton technique[20] with a smearing parameter of 0.2 eV is also used. The total energy is converged numerically to less than $1\times10^{-6}$ eV/unit. After structural optimization, calculated forces are converged to less than 0.01eV/Å. As mentioned in our previous work[21], the Vinet[22] equation of state is used in this work to obtain the equilibrium volume ($\Omega_0$), and the total energy (E). Here in opposition to our preliminary work[23] we have performed the calculations with and without the spin-orbit coupling (SOC) and we will show the importance of the spin-orbit coupling for the physical properties related to the electronic structure.

## 3. Results and discussion

### 3.1 Structural properties

In the present calculations, the lattice constants and formation enthalpies of the three compounds have been calculated with and without SOC, and listed in Table 1. For $Tl_5Te_3$, the present calculated lattice constants with and without SOC are in good agreement with the experimental data[9]. The slight difference is due to the Generalized Gradient Approximation (GGA) used in this work since it is well known that the GGA overestimates the lattice constants or the equilibrium volume[24]. The lattice constants of $BiTl_9Te_6$ and $SbTl_9Te_6$ (see Table 1) are in good agreement with the experimental data[9] as well. From Table 1, it can also be seen that the lattice constants of the three compounds with SOC are slightly closer to the experimental data than the results without SOC. The formation enthalpies of the three compounds are also listed in Table 1, and the value of $Tl_5Te_3$ is in good agreement with the experimental values [25-27]. The results also show that both $BiTl_9Te_6$ and $SbTl_9Te_6$ have a lower formation enthalpy than that of $Tl_5Te_3$ which indicates that the ternary compounds $BiTl_9Te_6$ and $SbTl_9Te_6$ are more stable than $Tl_5Te_3$.

3.2 Electronic structure

We represent in Fig.1 the total density of states of the three compounds with and without SOC. As shown in Fig.1, whereas $Tl_5Te_3$ is metallic, the substituted compounds become semiconducting which *a priori* is favorable for thermoelectricity[1]. Concerning the effect of the SOC, for the substituted compounds, the main differences concern the width of the conduction bands which is larger with the SOC and the width of the gap which is smaller with the SOC. The SOC calculated band gap of $BiTl_9Te_6$ is equal to 0.26 eV (0.58 eV without SOC) and the one of $SbTl_9Te_6$ is equal to 0.37 eV (0.54 eV without SOC). The calculated band gaps without SOC are coherent with the experimental results ($BiTl_9Te_6$: $E_g \geq 0.4$ eV[10]; $SbTl_9Te_6$ : Eg ≈ 0.5 eV[11]) whereas the ones obtained with SOC are smaller than the experimental values which is the usual behavior expected from DFT-GGA calculations[24]. Also, it should be noted that at the Γ point, without SOC, the bandgap is only slightly larger than the indirect bandgap (0.62 eV), whereas with SOC the bandgap is equal to 0.49 eV, which is roughly 2 times larger than the indirect bandgap. Thus, the inclusion of the SOC is crucial to obtain a good description of the band structure at the vicinity of the Fermi level, as it has also been shown in other semiconductors containing heavy elements as like $Bi_2Te_3$[28,29]. In that case, as in our case, the effect of the SOC is double: it broadens the lowest conduction bands and hence reduces the size of the bandgap and it changes the position of the valence and conduction bands extrema as we will see later.

We show the partial density of states and the band structure (calculated with the SOC) of $Tl_5Te_3$ and $BiTl_9Te_6$ in Fig. 2 and Fig. 3, to understand the nature of the bonding mechanism (the electronic properties of $SbTl_9Te_6$ are similar to those of $SbTl_9Te_6$ and are therefore not shown). Fig. 2(a) shows the calculated electronic density of states of $Tl_5Te_3$, while Fig. 2(b) shows the band structure of $Tl_5Te_3$ along the high symmetry directions M→Γ→X→P→N→Γ. There is a narrow band gap observed above the Fermi level, which is consistent with the metallic behavior of $Tl_5Te_3$. Our results agree qualitatively with those of Nordell and Miller who performed tight-binding calculations using the extended Hückel method[5]. We notably confirm the presence of one hole pocket around the Γ point, of one electron pocket around the point N and that the Fermi level intersects degenerate bands at point P (it is worth noting that without the SOC this degeneracy is not present). As noticed by Nordell and Miller[5], this leaves the possibility of an electronic instability that could be related to the superconductivity observed in $Tl_5Te_3$ and some of its alloys. We find that the 5d core levels are close to 10 eV at a lower energy than in XPS experiments[7]. But we confirm the different band assignations done by Lippens et al in XPS experiments[7] with the help of tight-binding calculations. Two broad features are observed at around -6 eV and -1 eV and are respectively due to essentially 6s levels of Tl atoms and 5p levels of Te atoms with some small contributions due to 6p levels of Tl. However, we find that the small contribution from the 6s levels of the Tl atoms is located at the Fermi level and not below.

The partial density of states and the band structure of $BiTl_9Te_6$ are shown in Fig. 3(a) and Fig. 3(b) respectively. At low energy, the DOS contribution between -7.0 eV and -4.0 eV is mainly due to the 6s states of Tl1 and Tl2. The region from -4.0 eV to the Fermi level as well as the conduction band is dominated by Tl1-6p, Tl2-6p, Te1-5p, Te2-5p, and Bi-6p states. From Fig. 3(b) it can be seen that the valence band maximum occurs along the ΓM direction whereas the conduction band minimum occurs along the ΓX direction. It is important to note that including SOC changes the position of the bands near the Fermi level. In the case of $BiTl_9Te_6$, if the main minimum of the conduction band is at the same position without and with SOC, the valence band maximum occurs along the NΓ direction without SOC and not along the MΓ direction. This difference must be of fundamental importance for the transport and thermoelectric properties of $BiTl_9Te_6$ since it is a p-type semiconductor. However, this does not change the following qualitative conclusion: $BiTl_9Te_6$ is an indirect narrow band gap semiconductor (like $SbTl_9Te_6$) with several valleys in the bandstructure which is *a priori* favorable for good thermoelectric properties. Indeed, it is well known that multivalley narrow gap semiconductors have favorable thermoelectric properties (like $Bi_2Te_3$[1,28]).

To visualize the nature of the bonds and to explain the charge transfer and the bonding properties of $Tl_5Te_3$ and $BiTl_9Te_6$ we have investigated the bonding charge density with SOC. The bonding charge density, also called the deformation charge density, is defined as the difference between the self-consistent charge density of the interacting atoms in the compound and a reference charge density constructed from the superposition of the non-interacting atomic charge density at the crystal sites. The distributions of the bonding charge densities in the (001) planes of Tl$_5$Te$_3$ and BiTl$_9$Te$_6$ compounds have been calculated to understand the alloying mechanism and are shown in Fig. 4 (a) and 4(b) respectively (SbTl$_9$Te$_6$ is very similar to BiTl$_9$Te$_6$ and is not shown)

In Fig. 4 (a), the bonding charge density of Tl$_5$Te$_3$ shows a depletion of the charge density at the Te sites together with a "horseshoe" shaped increase of the electronic density around the Te sites and a slight increase of the charge density at the Tl sites together with a depletion of the charge density around the Tl sites. It can thus be seen that the iono-covalent nature of the bonding between Tl and Te is dominant in the (001) plane: this is consistent with the PDOS plots in Fig. 2 (a) showing the importance of the Tl-6p and Te-5p hybridization. From Fig. 4 (b) the significant redistribution of the bonding charge between Bi and the Te atoms indicates the occurrence of a new strong iono-covalent bonding with the symmetry breaking of the "horseshoe" arrangement. This important redistribution of the electronic charge between the Bi and Te atoms is probably at the origin of the band gap appearing in the BiTl$_9$Te$_6$ (and SbTl$_9$Te$_6$) compounds.

3.3 Thermoelectric properties

In this section we determine the Seebeck coefficient at 300K in order to compare the thermoelectric properties of the different compounds. To determine the Seebeck coefficient (or the thermopower) α, we have used Mott's law[30] :

$$\frac{\alpha}{T} = -\frac{\pi^2 \kappa_B^2}{3e}\left(\frac{\partial \ln(\sigma(\varepsilon))}{\partial \varepsilon}\right)_{\varepsilon_F} \quad (1)$$

If we assume a negligible k-dependence of the group velocity close to the Fermi level and with the hypothesis of constant relaxation time, we obtain:

$$\frac{\alpha}{T} = -\frac{\pi^2 \kappa_B^2}{3e}\left(\frac{\partial \ln(n(\varepsilon))}{\partial \varepsilon}\right)_{\varepsilon_F} \quad (2)$$

where n(ε) is the electronic density of states.

This is a crude approximation but as discussed by Tobola and coworkers[31], in favorable cases, eq. (2) can give qualitative informations about the doping dependence of the thermopower.

To determine the thermopower at 300K of Tl$_5$Te$_3$ we simply applied Eq. (2) and therefore determine the logarithmic derivative of the electronic density of states at the Fermi level multiplied by -2.441 10$^{-2}$ (keeping the energy in eV). We obtain the relatively large value of -44 µV/K which is much larger than the experimental value (2 µV/K)[32]. Here, the disagreement is not surprising since experimentally this compound is not stoechiometric whereas our computer sample is perfectly stoechiometric. Indeed experimentally there exists an excess of thallium atoms accommodated on one of the two tellurium sites[33] which leads to a downshift of the Fermi level and hence changes the Seebeck coefficient . because from Fig.1 one can see that the Fermi level is located in a region in which the density of states changes very rapidly and thus a slight change in the approximations used in the calculations can induce a change of slope of the density of states at $E_f$ : we are probably at the limits of validity of the application of Mott's Law.

In the case of the doped samples, the Mott's law which is *a priori* only valid for metals can also be applied to determine the thermopower of semiconductors since the doped compounds are both degenerate semiconductors. This can be seen from the experimental reports in which metallic behaviours were reported for the electrical transport[10,11]. To determine the thermopower, we have used the rigid band model (RBM)[34] knowing the concentration of charge carriers and then we have applied Mott's law : this is a rough approximation but the aim is to check if this crude method can at least give some hints on the thermoelectric properties of the alloys. When the DOS is changing smoothly then the RBM is working. Conversely, in presence of some van Hove singularities or if the DOS is changing rapidly, the RBM does not work, as in the case of the pure $Tl_5Te_3$ compound. We can apply this method to the Bismuth and Antimony doped compound since no more than two bands contribute to the electronic transport and because of the absence of van Hove singularities. In addition we know that at 300K the hole concentrations are respectively 1.7 10$^{19}$ cm$^{-3}$ and 1.5 10$^{20}$ cm$^{-3}$ from experiments[10,11]. So we apply the rigid band model and lower the Fermi level on the DOS of Fig.1 until we reach the correct carrier concentration. Finally we apply Eq. (2) and obtain a thermopower at 300K of 163 µV/K for $BiTl_9Te_6$ and of 200 µV/K for $SbTl_9Te_6$. This is coherent with the experimental values of 260 µV/K found for $BiTl_9Te_6$ and of 120 µV/K for $SbTl_9Te_6$[10,11]. However, we do not find the experimental tendency and therefore some improvements are necessary to reproduce quantitatively the experiments. We note that Yamanaka et al. found a thermopower very close to our results (160 µV/K)[35] but unfortunately the charge carriers in their samples are not known (it is worth noting that including the SOC gives a better agreement with experiments. Indeed without SOC, we find a thermopower at 300K of 137 µV/K for $BiTl_9Te_6$ and of 82 µV/K for $SbTl_9Te_6$).

We can expect that using post-DFT methods such as GW methods to obtain a more quantitative band structure would help to obtain a better quantitative agreement, but these techniques need huge calculation resources, especially for the case of heavy atoms requiring the SOC like in the present case.

Finally even though we have used a rough approximation, we find clearly that the Bi and Sb doped compounds have promising thermoelectric properties, in agreement with experimental results. This is probably related to the multivalley character of these narrow gap semiconducting compounds as mentioned earlier.

### 3.4 Elastic properties

In order to shed some light on the mechanical properties of $Tl_5Te_3$, $BiTl_9Te_6$, and $SbTl_9Te_6$, the elastic constants of the three compounds have been calculated with SOC in the present work and are listed in Table 2.

For a tetragonal system, the mechanical stability criteria are given by $C_{11}>0$, $C_{33}>0$, $C_{44}>0$, $C_{66}>0$, $(C_{11}-C_{12})>0$, $(C_{11}+C_{33}-2C_{13})>0$, and $[2(C_{11}+C_{12})+C_{33}+4C_{13}]>0$. The calculated elastic constants of $Tl_5Te_3$, $BiTl_9Te_6$, and $SbTl_9Te_6$ satisfy these stability conditions. In the present calculations, $C_{11}>C_{33}$, indicating that the bonding strength along the [100] and [010] direction is stronger than that of the bonding along the [001] direction. It is worth noting that $C_{44}$ is almost equal to $C_{66}$ for $Tl_5Te_3$, $BiTl_9Te_6$, and $SbTl_9Te_6$ compounds. This suggests that the [100](010) shear is similar to the [100](001) shear for these compounds. The shear elastic anisotropy factor[36] can be estimated as $A=2C_{66}/(C_{11}-C_{12})$. If A is equal to one, no anisotropy exists. From Table 2, the shear elastic anisotropy factor of the $Tl_5Te_3$, $BiTl_9Te_6$ and $SbTl_9Te_6$ phase is almost equal to one indicating that the shear elastic properties of the (001) plane are nearly independent of the shear direction.

The bulk modulus, shear modulus, Young's modulus and Poisson's ratio have been estimated from the calculated single crystal elastic constants, and are given in Table 2. The bulk moduli of the $Tl_5Te_3$, $BiTl_9Te_6$, and $SbTl_9Te_6$ compounds estimated from the single crystal elastic constants and the results in Table 2. show that the bulk modulus of $BiTl_9Te_6$ and $SbTl_9Te_6$ is smaller than the one of $Tl_5Te_3$, It can also be noted that the bulk moduli, shear moduli, and Young's moduli of $Tl_5Te_3$,

BiTl$_9$Te$_6$, and SbTl$_9$Te$_6$ are relatively close to each other.

It is well known that the Poisson coefficient, ν, is close to 1/3 in metallic compounds, to 1/4 for ionic materials and is small in covalent materials (typically 0.1) [37]. Here, we find that ν = 0.28 in the metallic Tl$_5$Te$_3$ and is smaller in the semiconducting BiTl$_9$Te$_6$ and SbTl$_9$Te$_6$ (ν = 0.22) which agrees well with the tendencies reported above. Obviously the mechanical properties of the studied materials are related to their bonding characteristics. The relatively large value of the Poisson coefficient in the two semiconducting compounds suggests that their bondings have a relatively strong ionic character, in good agreement with our discussion of the charge density map (Fig. 4b).

The ratio between the bulk and the shear modulus, B/G, has been proposed by Pugh[38] to predict brittle or ductile behavior of materials. According to the Pugh criterion, a high B/G value indicates a tendency for ductility. If B/G>1.75, then ductile behavior is predicted, otherwise the material behaves in a brittle manner. This ratio for BiTl$_9$Te$_6$ and SbTl$_9$Te$_6$ is close to 1.75, while it is larger than 1.75 for the metallic Tl$_5$Te$_3$. These results suggest that the semiconducting BiTl$_9$Te$_6$ and SbTl$_9$Te$_6$ are slightly prone to brittleness while Tl$_5$Te$_3$ is prone to ductility. This finding is clearly related to the change in the electronic structure and of the bonding character due to substitution of Te by Bi or Sb. Indeed, metallic compounds are known to be usually ductile (with some exceptions such as iridium)[37,39] whereas ionic and more specifically covalent compounds are generally somewhat brittle[37]. The fact that BiTl$_9$Te$_6$ and SbTl$_9$Te$_6$ are close to the ductile-brittle limit confirms the presence of some ionic character in their bonding. It is interesting to see that the Bi and Sb substitution in metallic Tl$_5$Te$_3$ not only changes the band structure and moves the Fermi level in the band gap (making BiTl$_9$Te$_6$ and SbTl$_9$Te$_6$ semiconducting) but also changes the mechanical properties from ductile to brittle behavior. This shows a nice example of the correlation between the electronic properties, the nature of bonding and the mechanical properties.

## 3.5 Thermodynamic properties

For the three compounds we determine the Debye temperature ($\Theta_D$) from the averaged sound velocity ($v_D$) using the following equation:

$$\Theta_D = \frac{h}{k_B}\left(\frac{3}{4\pi V_a}\right)^{1/3} v_D \tag{3}$$

where $h$ and $k_B$ are, respectively, Planck's and Boltzmann's constants and $V_a$ is the atomic volume. The average sound velocity in polycrystalline systems, $v_m$, are evaluated by

$$\frac{1}{v_D^3} = \frac{1}{3}\left(\frac{1}{v_l^3} + \frac{2}{v_t^3}\right) \tag{4}$$

where $v_l$ and $v_t$ are the mean longitudinal and transverse sound velocities, which can be related by the shear and bulk moduli:

$$v_l = \left(\frac{3B+4G}{3\rho}\right)^{1/2} \quad \text{and} \quad v_t = \left(\frac{G}{\rho}\right)^{1/2} \tag{5}$$

The calculated longitudinal, transverse and average sound velocities and Debye temperature of $Tl_5Te_3$, $BiTl_9Te_6$, and $SbTl_9Te_6$ compounds have been calculated and listed in Table 3. The low values found are obviously mainly related to the large mass of the atoms composing these materials. The calculated average sound velocities and Debye temperatures of the three compounds are very similar, and the values for $BiTl_9Te_6$ are close to the experimental data[11]. As far as we know, there are no experimental values available for $Tl_5Te_3$ and $SbTl_9Te_6$ and we are not aware of other theoretical data for these quantities.

## 4. Conclusion

We have presented results of first-principles calculations for the phase stability, the electronic structure, the thermoelectric, elastic and thermodynamic properties of $Tl_5Te_3$, $BiTl_9Te_6$, and $SbTl_9Te_6$ compounds. We find that including the spin-orbit coupling is mandatory since its influence is quite significant as for other telluride compounds. The calculated results accurately predict the stable phase of $Tl_5Te_3$ within the $Cr_5B_3$-type structure. The calculated lattice constants are in good agreement with experimental data. The density of states and bonding charge densities of the three compounds have been calculated and indicate that iono-covalent bonding is dominant in the (001) plane. We have calculated the thermopower of the doped samples using very simple and rough approximations and found a good qualitative agreement. The elastic constants and elastic moduli of $Tl_5Te_3$, $BiTl_9Te_6$, and $SbTl_9Te_6$ have been calculated in this work, but as far as we know, there are no experimental data available for these quantities. Experiments are expected to validate the present numerical data. The observed change from ductile to brittle behavior with Bi or Sb alloying is suggested to be related to the change of electronic structure from metallic behavior in pure $Tl_5Te_3$ to semiconducting behavior after alloying. The calculated Debye temperature and average sound velocity of $BiTl_9Te_6$ are in agreement with experimental data. The present results give hints for the design of materials based on $Tl_5Te_3$ and should be used to stimulate future experimental and theoretical work.

**Acknowledgments:** X. T. thanks the "Fondation d'enterprise E.A.D.S." and the Scientific Research Foundation of Guangxi University (Grant No. XBZ100022) for the support of his research. P. J. thanks the computer center hpc@lr in Montpellier and Microsoft for their support.

Table Captions

Table 1 Calculated lattice constants and formation enthalpies of $Tl_5Te_3$, $BiTl_9Te_6$ and $SbTl_9Te_6$ compounds

Table 2 Calculated elastic constants, bulk modulus, shear modulus, Young's modulus, Poisson's ratio and B/G of $Tl_5Te_3$, $BiTl_9Te_6$ and $SbTl_9Te_6$

Table 3 Calculated longitudinal, transverse, average sound velocities, and Debye temperature of $Tl_5Te_3$, $BiTl_9Te_6$ and $SbTl_9Te_6$

Figure Captions

Fig. 1 Total density of states of $Tl_5Te_3$, $BiTl_9Te_6$ and $SbTl_9Te_6$

Fig. 2 (a) Partial density of states and (b) band structure of $Tl_5Te_3$

Fig. 3 (a) Partial density of states and (b) band structure of $BiTl_9Te_6$

Fig. 4 Bonding charge density in the (001) plane of (a) $Tl_5Te_3$ and (b) $BiTl_9Te_6$

Table 1

| Phase | Lattice parameters (Å) | | Formation enthalpy (ΔH) eV/atom | Remarks |
|---|---|---|---|---|
| | a | c | | |
| Tl$_5$Te$_3$ | 8.994 | 13.162 | -0.279 | without SOC |
| | 8.991 | 13.094 | -0.243 | with SOC |
| | 8.930[9] | 12.589[9] | -0.224[24] | Experiments |
| | | | -0.278[25] | |
| | | | -0.280[26] | |
| BiTl$_9$Te$_6$ | 8.980 | 13.499 | -0.308 | without SOC |
| | 8.952 | 13.458 | -0.261 | with SOC |
| | 8.859[9] | 13.458[9] | | Experiments |
| SbTl$_9$Te$_6$ | 8.948 | 13.480 | -0.286 | without SOC |
| | 8.924 | 13.426 | -0.246 | with SOC |
| | 8.847[9] | 13.024[9] | | Experiments |

Table 2

|  | $Tl_5Te_3$ | $BiTl_9Te_6$ | $SbTl_9Te_6$ |
|---|---|---|---|
| $C_{11}$ (GPa) | 41.16 | 39.08 | 38.25 |
| $C_{12}$ (GPa) | 13.79 | 13.86 | 12.88 |
| $C_{33}$ (GPa) | 30.10 | 29.50 | 26.36 |
| $C_{13}$ (GPa) | 21.52 | 12.75 | 12.14 |
| $C_{44}$ (GPa) | 12.91 | 11.91 | 12.18 |
| $C_{66}$ (GPa) | 12.27 | 12.20 | 12.01 |
| Bulk Modulus B (GPa) | 25.08 | 20.50 | 19.37 |
| Shear Modulus G (GPa) | 10.50 | 11.70 | 11.53 |
| Young's Modulus E (GPa) | 27.64 | 29.49 | 28.86 |
| Poisson's ratio v | 0.28 | 0.22 | 0.22 |
| elastic anisotropy A | 0.896 | 0.967 | 0.947 |
| B/G | 2.39 | 1.75 | 1.68 |

Table 3

|  | $v_l$ [m/s] | $v_t$ [m/s] | $v_D$ [m/s] | $\Theta$ [K] |
|---|---|---|---|---|
| $Tl_5Te_3$ | 2105 | 1091 | 1221 | 113 |
| $BiTl_9Te_6$ | 2041 | 1161 | 1292 | 119 |
|  |  |  | 1300[11] | 107[11] |
| $SbTl_9Te_6$ | 2025 | 1167 | 1295 | 120 |

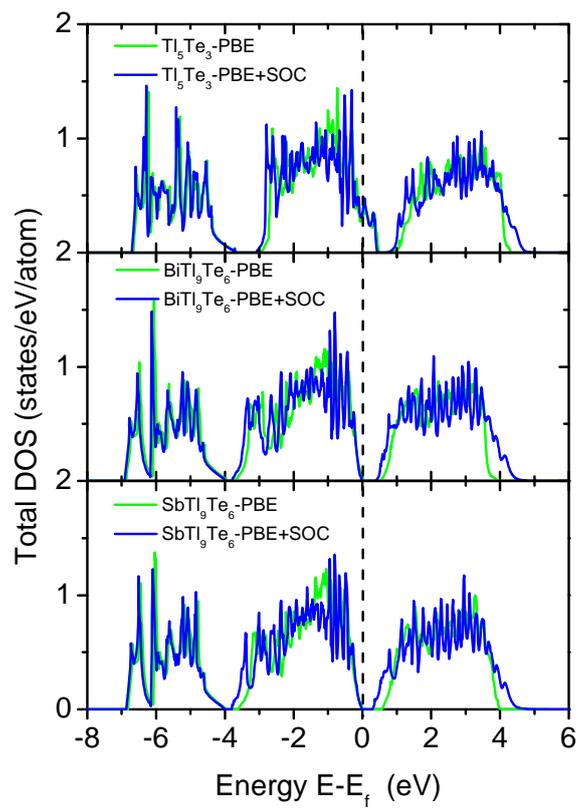

Fig. 1

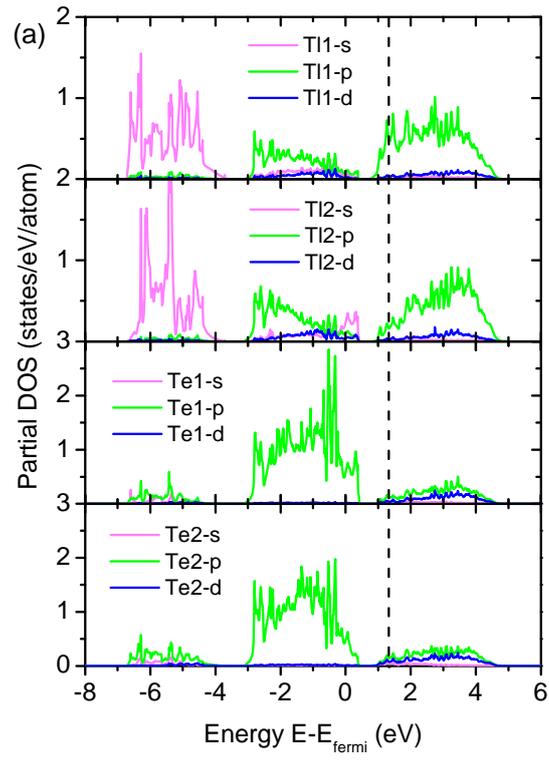

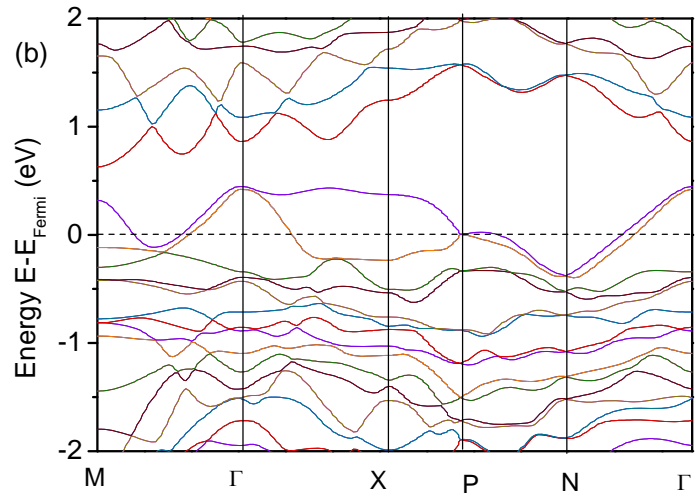

Fig. 2

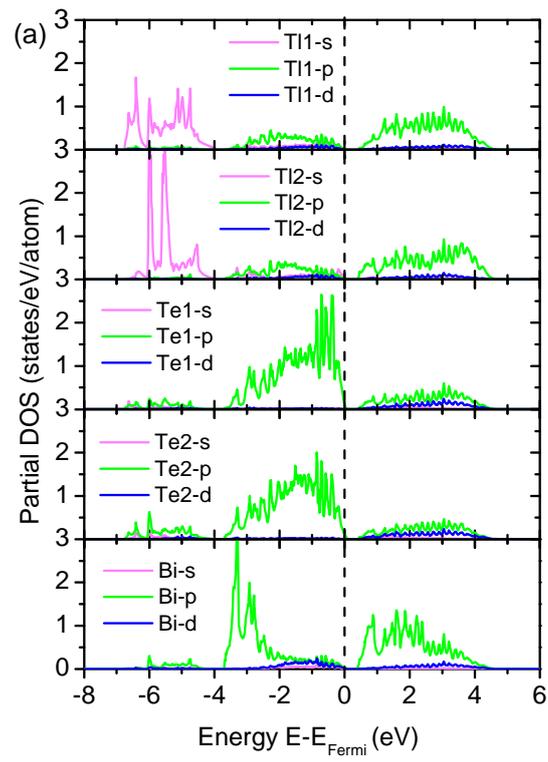

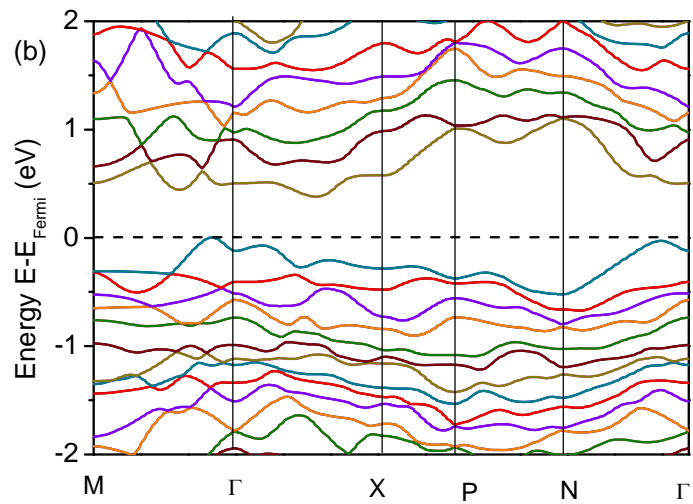

Fig. 3

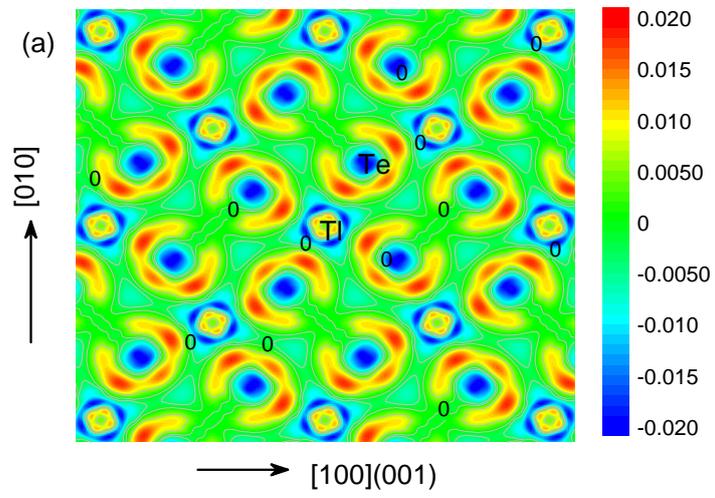

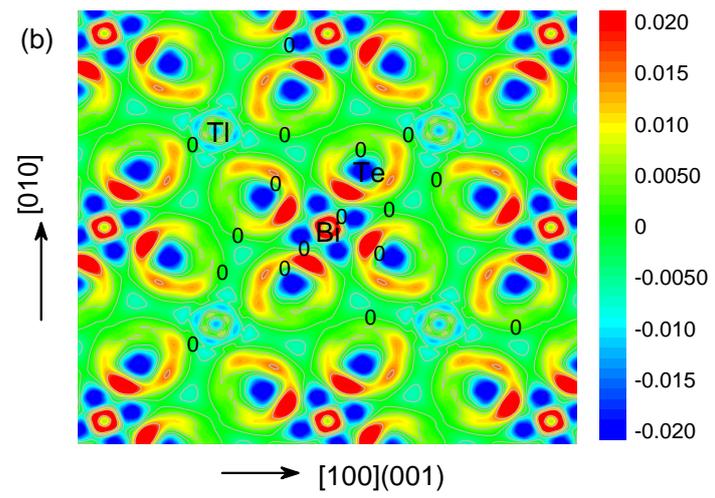

Fig. 4